# Implementation of Firm-Dispatchable Generation in South Africa

Stephen R. Clark *, Craig McGregor

Department of Mechanical and Mechatronic Engineering, Stellenbosch University, Stellenbosch, South Africa *(corresponding author: Sclark@sun.ac.za)

**Abstract**

South Africa is currently facing a critical situation in its power generation landscape, which is plagued by frequent power outages and the need to move from fossil fuels to renewable energy sources. This period emphasizes the importance of having firm-dispatchable power to balance out the intermittent nature of wind and solar energy sources. The paper proposes to repurpose old coal-fired power plants to generate firm-dispatchable energy in line with the principles of a Just Transition. Eskom's coal plants are approaching the end of their economic life, and their declining energy availability factor is becoming a challenge in meeting the country's energy needs. The study suggests that a comprehensive strategy that integrates wind, solar, and firm-dispatchable power can be cost-effective and reliable compared to the traditional coal-based approach or the nuclear alternative. The study emphasizes the necessity of a 25-year plan that would invest in flexible and modular dispatchable generation. It also highlights the strategic location of this generating capacity, including repurposing decommissioned coal plant sites. The proposed model integrates private investment, adheres to established best practices, and emphasizes adaptability to changing demand dynamics. The study provides a roadmap for enabling firm-dispatchable capacity for South Africa's energy transition, emphasizing economic prudence, environmental sustainability, and alignment with the principles of the Just Transition program.

**Introduction**

South Africa is shifting from a power generation system that relies on fossil fuels to one based on renewable energy sources such as wind and solar power. This transition is motivated by the need to decrease greenhouse gas emissions and is also becoming a more cost-effective option (Creamer, 2023). Furthermore, the government has stated that this transition should be carried out within the framework of the Just Transition program where the transition is done in a manner that preserves employment in affected areas (Govt of South Africa, 2023).

Wind and solar energy sources are known for their variability, so there must be a provision of firm-dispatchable power to meet the demand when these sources cannot. A program for firm-dispatchable power must be implemented to successfully transition to a renewable energy system that relies on variable sources (Clark & McGregor, 2024). South Africa's current generation system is not meeting its needs, and the situation will likely worsen as base load facilities continue to age and must be replaced. This review summarises the current generation system and considers the implementation of the necessary firm-dispatchable generation system. Repurposing current baseload coal generation plants as potential locations for developing firm-dispatchable generation facilities is proposed.

**Background**

Eskom is the sole electricity provider in South Africa, generating most of its power from fifteen coal-fired plants that operate constantly (Eskom, n.d.). As of 2023, the average age of these plants was 36 years. According to international experience, coal-fired power plants usually have an average maximum economic lifespan of 46 years (Cui et al., 2019). Eskom plans their plants to last for 50 years, so many of them are nearing the end of their useful lives and will soon need to be decommissioned (Department of Energy, 2011). Of the 39.8 GW total of installed coal-fired plants, Eskom anticipates decommissioning 28 GW in the next 15 years and nearly 35 GW in the next 25 years, as illustrated in Figure 1 (Futuregrowth Asset Management, 2021).

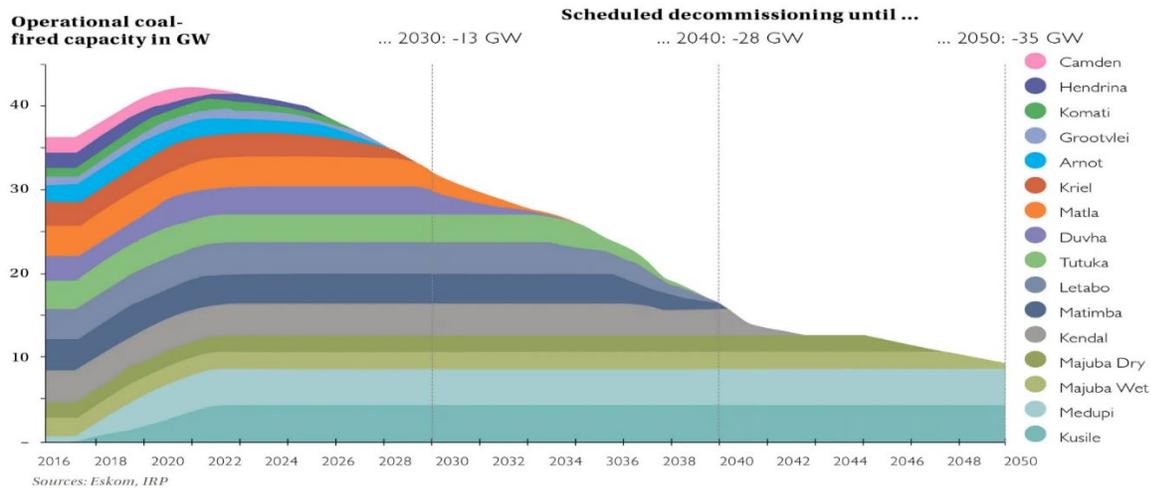

*Figure 1 - Eskom decommissioning plan* (Futuregrowth Asset Management, 2021)

It has been observed globally that thermal plants have a decreasing average capacity factor as they get older. This factor is called the Energy Availability Factor (EAF) in South Africa. In the last two decades, the EAF for the South African generation fleet has decreased from 80% to 53%, as indicated in Figure 2 (Pierce & le Roux, 2023). The Integrated Resource Plans (IRPs) created since 2010 assumed an increase the EAF of these plants to meet growing energy demands (Department of Energy, 2019). However, experience has shown that this is unlikely to happen, and the fleet's performance will continue to decline until each plant becomes financially unviable. A study conducted in the USA by the National Energy Technology Laboratory (NETL), which analysed the performance of over 300 coal plants in the USA, has shown that the performance of ageing plants declines just as experienced by the South African fleet, as illustrated in Figure 3 (Kern, 2015).

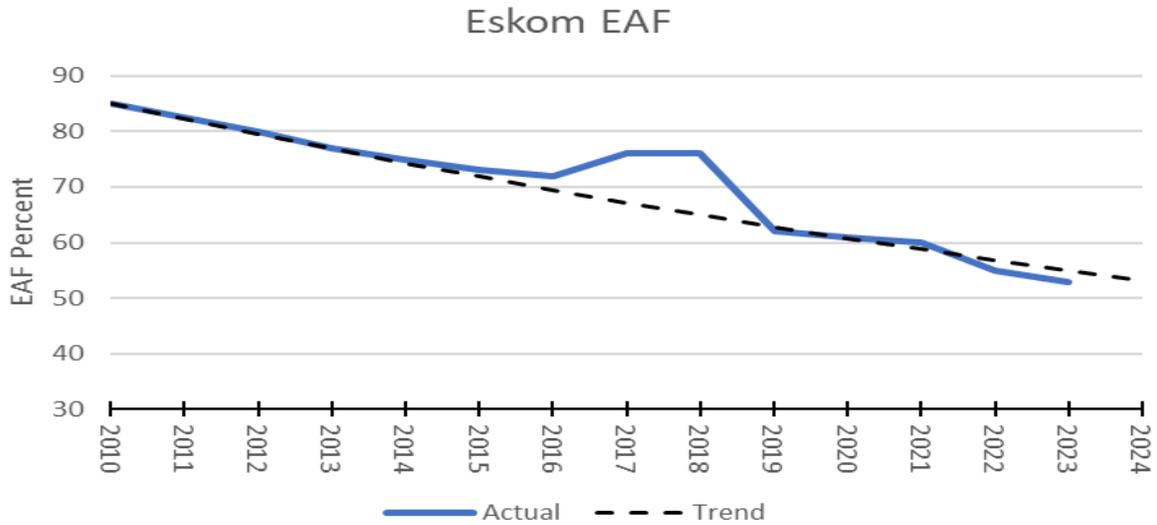

*Figure 2 - Eskom fleet EAF history – data* (Pierce & le Roux, 2023)

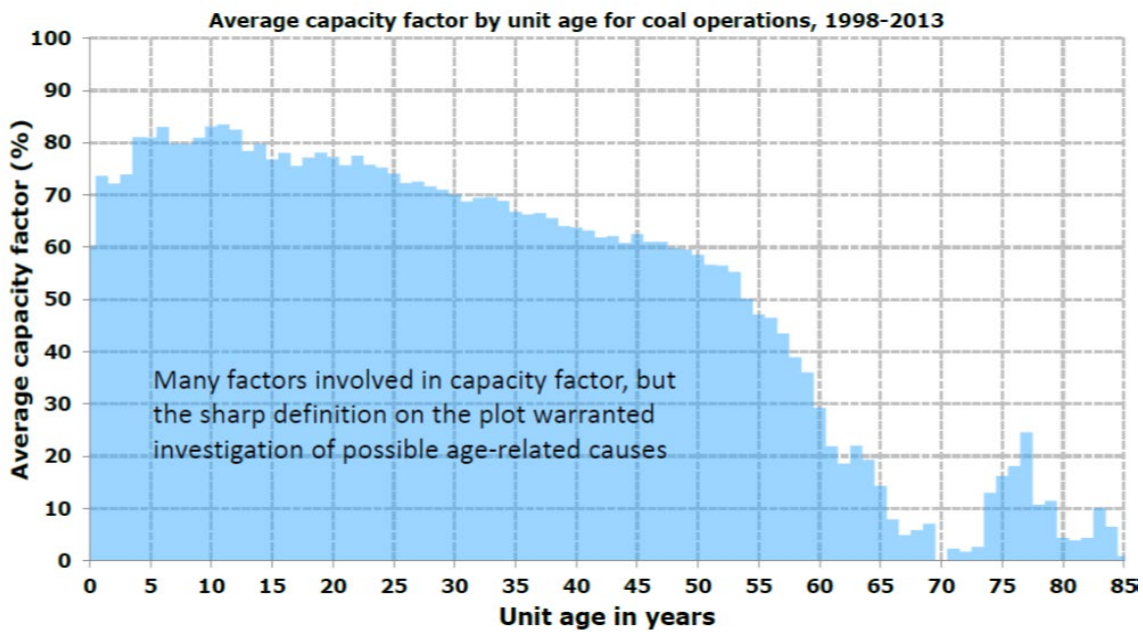

*Figure 3 - USA study on coal plant CF with age* (Kern, 2015)

It is important to note that older power plants face numerous challenges in addition to maintaining their performance. They also produce harmful pollutants during coal combustion, which can cause serious health problems to people exposed to power station emissions (Bloomberg, 2023). To address this issue, significant modifications are required to reduce pollution, which can be expensive. However, due to the need for power generation and the high cost of these modifications, most of the Eskom plants have been granted exemptions from installing the required pollution control (Bloomberg, 2023). In the latest proposed IRP, more exemptions have been requested (Department of Mineral Resources and Energy,

2024). These license-to-operate concerns may ultimately lead to some non-compliant plants being forced to shut down before they reach their expected 50-year lifespan.

South Africa has pledged to reduce its greenhouse gas emissions in accordance with the Paris Accord, as have most other countries worldwide (UNFCCC, 2015). As part of the South African government's plan, the country has committed to phasing out coal-fuelled power plants to meet these commitments (UNFCCC, n.d.). In addition to the economic reasons for discontinuing the use of these aging facilities, meeting these commitments will also put pressure on the country to decommission these plants and replace them with low or zero greenhouse gas-emitting sources of electricity. While recognising the need to close out the large coal fuelled generation plants, the government is cognisant of the effect that these plants closure will have on local employment (Govt of South Africa, 2023). These plants are major sources of employment in their respective locations, with jobs in the plants plus mining and related jobs associated with fuelling these plants with local coal. To minimise the impact on these communities, the government has mandated that the transition be conducted within the framework of a "Just Transition" where jobs are maintained as much as possible and where-ever possible new jobs created in those locations. This must be a part of the planning for any renewable based system.

One of the major questions that is always present in any forecast for power generation requirements is the forecasting of the demand profile. In 1990, a study on electricity forecasting in the USA stated (Hobbs & Maheshwari, 1990):

> *"Planning decisions are severely affected by uncertainties in demand. This fact has two implications: accurate forecasts of demand are worth more than forecasts of other parameters, and the expected cost of ignoring demand uncertainties exceeds the cost of disregarding other sources of risk."*

In the past, it was believed that the demand for electricity would grow at a rate roughly equal to the growth in GDP in the area being served (Sharma, Smeets & Tryggestad, 2019). However, over the last 15 years, the power generation from the central power provider in South Africa has been decreasing steadily. On average, demand has been decreasing by 0.5% per year, and this rate of decrease appears to be increasing, as illustrated in Figure 4 (Pierce & le Roux, 2023). This trend is likely due to several factors, including the adoption of energy-efficient technologies such as LED lighting, the conversion of heavy energy users to less intense energy operations, and the rise in self-generation (such as behind-the-meter solar installations). While some argue that the rapid electrification of major power uses, such as the growth of the electric vehicle market, will reverse this trend and lead to large increases in power demand, there is a significant uncertainty about the future demand (Merven, 2023). Therefore, any plan to address future electricity demand must be flexible enough to adapt to uncertainty and changes in demand growth. Building baseload generation plants to meet unknown demand is challenging because of the time required to construct these plants and the economies of scale associated the size of the plants. This can result in either a capacity under build, which leads to unserved demand, or an over build, which results in higher costs.

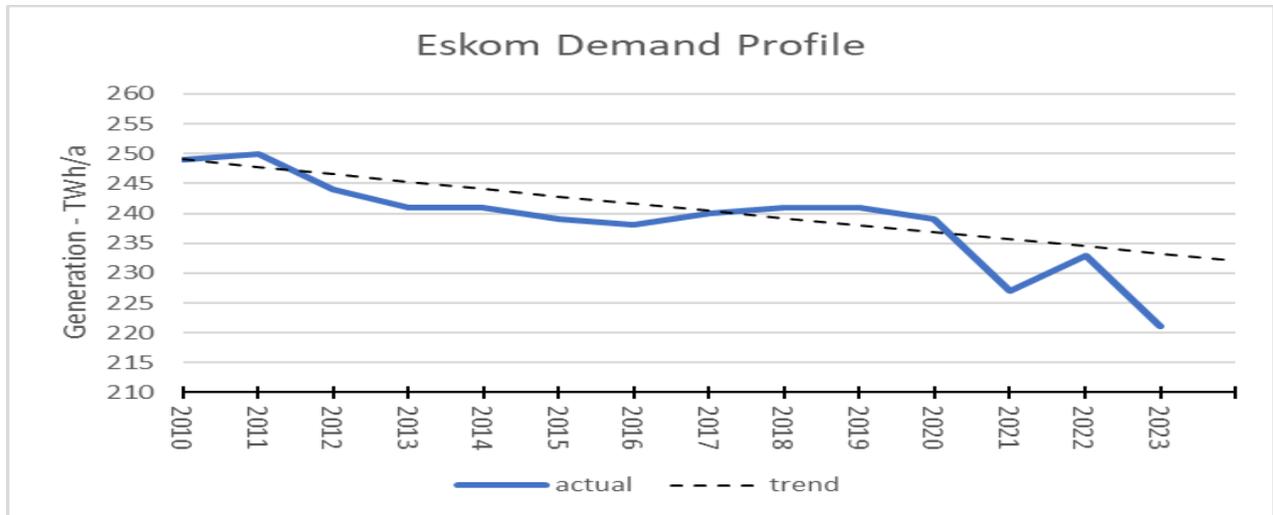

*Figure 4 - Eskom annual generation 2010-2023 – data* (Pierce & le Roux, 2023)

In June 2023, the South African Presidential Climate Commission (PCC) released its recommendations on South Africa's electricity system. The report suggested that a least-cost approach to generating electricity would require a significant increase in the construction of renewable energy sources supported by battery storage and *gas peakers* (Creamer, 2023). This transition to renewable energy sources is also being embraced globally, and many countries are rapidly increasing their wind and solar generation capacity. Numerous studies conducted by the CSIR, and other institutions, have corroborated the findings of the PCC report, which advocates for a least-cost system based on renewable energy sources (Bischof-Niemz, 2017; Roy, Sinha & Shah, 2020). However, there is a subtle yet significant difference in conclusion: the requirement is for *firm-dispatchable* power instead of *peaking* power (Clark & McGregor, 2024).

These studies have shown that even with the lowest cost generation system, it is still essential to have installed firm-dispatchable generation capacity to meet the system's needs during times when wind and solar power are unable to meet demand. This period of low renewable resources could happen for several days each year, and in some years, it could be for periods of several weeks (Clark, Van Niekerk & Petrie, 2020). This requirement was tested in models of the South Africa grid and models of the Texas and UK grids, which all showed similar results (Clark & McGregor, 2024). It is not a replacement for base load generation but would be utilised less than 10% of the time and must always be available. Operating baseload plants with this low utilisation rate would be expensive due to the high capital cost of thermal plants (Bergh & Delarue, 2015). International experience has shown that using the baseload plants in cycling mode reduces their expected lifespan by adding fatigue stress to elements of the plant that are not designed for this mode of operation (Kumar et al., 2012; Nichols, 2016).

**Size, Cost and Timing**

As mentioned earlier and illustrated in past analyses, it is necessary to have firm-dispatchable power generation available to fully support the portion of the total power generation that is being supplied by variable sources such as wind and solar. In other words, this firm-dispatchable power generation must be installed to completely substitute the volume of base load power that is being decommissioned, and it must be done at the same pace as the decommissioning process (Clark & McGregor, 2024). There is also a need to provide firm-dispatchable generation to complement the variable sources of generation (wind and solar) that have already been installed on the grid.

The current demand from the Eskom grid is 222 TWh per year, with a peak demand of 35 GW and an average of 25 GW over the year. Table 1 shows the sources of generation in the current system (Pierce & le Roux, 2023). As mentioned earlier, coal is the primary generation source, but the capacity factor of the coal fleet is low.

*Table 1 - Eskom generation fleet 2022 – data* (Pierce & le Roux, 2023)

| Technology | Installed Cap - GW | Generation - TWh | Percent of total | Capacity Factor - % |
|---|---|---|---|---|
| Coal | 39.8 | 176.6 | 80.1 | 50.7 |
| Nuclear | 1.9 | 10.1 | 4.6 | 60.7 |
| Hydro + pumped | 3.3 | 14.0 | 6.4 | 48.4 |
| Wind | 3.4 | 9.7 | 4.3 | 32.6 |
| Solar (PV+CSP) | 2.8 | 6.5 | 2.8 | 26.5 |
| Dispatchable | 3.4 | 3.6 | 1.6 | 12.1 |

Fifteen coal-fired power plants are currently installed with a total rated electricity capacity of 39.8 GW. The power plants have different capacities, ranging from 990 MW to 4.8 GW, with an average size of 2.6 GW (Eskom, n.d.) The fleet's overall age is 36 years from the commissioning date. However, two recently built plants contribute nearly a quarter of the installed capacity, 9.6 GW out of the total 39.8 GW capacity (Eskom, n.d.). Without the two new plants, the rest of the fleet has an age range from 22 to 57 years and an average age of 41. Eskom has planned a decommissioning schedule as indicated above in Figure 1.

The output for the nuclear plant located at Koeberg has dropped from approximately 85% in 2016 to the current EAF in the range of 65%, according to reporting from Eskom. This EAF decrease for Koeberg is shown in Figure 5 (Eskom, 2016, 2017, 2018, 2019, 2020, 2021, 2022, 2023). This plant has recently undergone a major refurbishment and will remain operational until 2044. It is expected that this plant will have reached its economic life at that time and will be decommissioned near the end of the planning period.

Hydropower facilities generally have a longer life expectancy than thermal power plants (US Dept. of Energy, 2003). The 0.6 GW of hydro generation and 2.7 GW of pumped storage should be available throughout the current planning term until 2050.

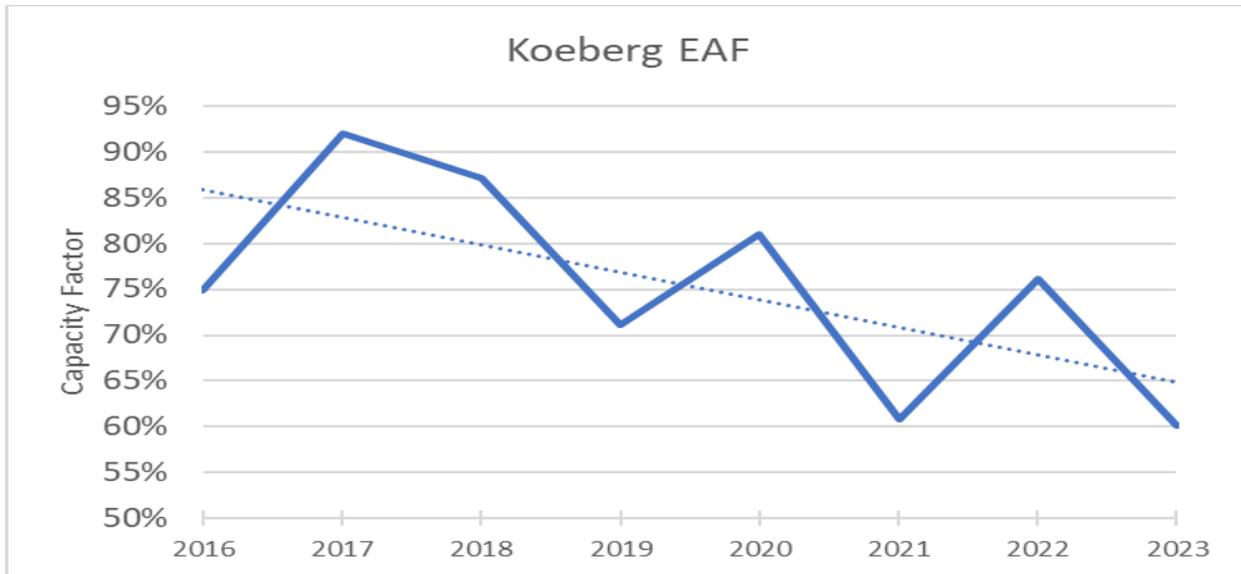

*Figure 5 - Koeberg output 2016-2023 – data (Eskom, 2016, 2017, 2018, 2019, 2020, 2021, 2022, 2023)*

Assuming the 35 GW of current baseload generation needs to be replaced with equivalent plants, this would mean constructing 25 GW of coal plants or nuclear (with an assumed 70% EAF) over the next twenty-five years. That is, on average, a capacity of 1000 MW is added annually. This assumes that the demand profile remains the same but with enough generation installed to eliminate load shedding. According to the US EIA capital cost estimates, shown in Figure 6, and detailed in Table 2, this would require an investment of USD 171 billion for these new coal power plants (US EIA, 2023). For nuclear, the capital cost would be slightly more, at $185 billion USD. However, the expected timing to build new baseload plants and the experience from the last two plants built in South Africa suggest a construction time of over 14 years, which means that this program is years behind what would be required (World Bank, 2023). This is a completely unrealistic scenario.

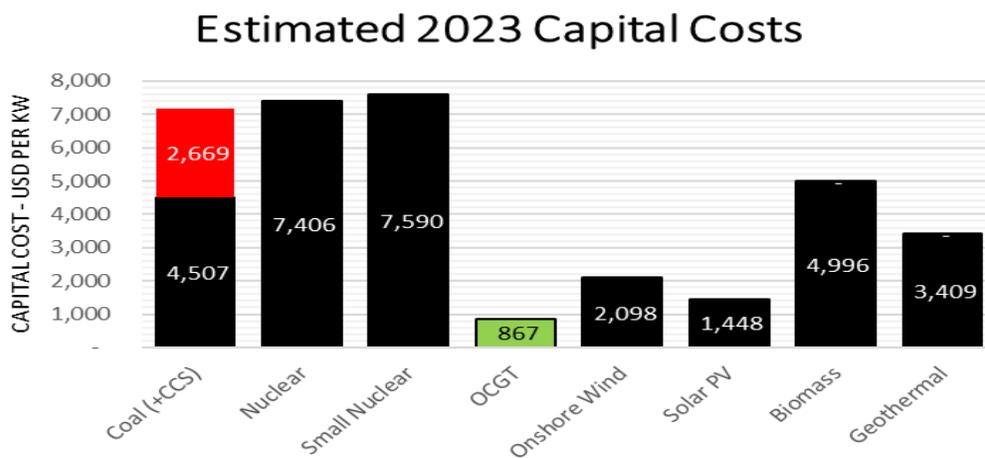

*Figure 6 – Power plant capital cost by technology – data (US EIA, 2023)*

*Table 2 - Costs to replace Eskom coal fleet*

| Technology | Unit Cost USD/kW | New Capacity GW | Total Cost Billion USD | Annual Cost Billion USD | Cost per MWh USD / MWh | Cost per kWh ZAR (19 ZAR/USD) |
|---|---|---|---|---|---|---|
| Base Load System | | | | | | |
| Coal | 6876 | 25 | 171 | 6.8 | 31 | 0.58 |
| Nuclear | 7406 | 25 | 185 | 7.4 | 33 | 0.63 |
| Renewable Based System | | | | | | |
| Wind | 2098 | 49 | 102 | 4.1 | | |
| Solar PV | 1448 | 14 | 20 | 0.8 | | |
| BESS | 400 (1) | 24 (2) | 10 | 0.4 | | |
| OGCT | 867 | 15 | 13 | 0.5 | | |
| Total Renewable | | | 145 | 5.8 | 26 | 0.50 |

(1) BESS costs are in USD per kWh of storage
(2) BESS capacity is in GWh of storage

The previous "business as usual" scenarios can be compared with a renewable-based generation system that mostly relies on wind and solar power. At the end of the 25-year planning period, this renewable system is expected to retain 12.9 GW of baseload, while wind and solar will replace all the decommissioned generation. The remaining baseload generation will come from the newly built coal-fuelled plants (Medupi and Kusile) and hydropower, which only need to be replaced beyond the end of the planning period. Based on the current demand profile, the model indicates that 49 GW of new wind and 14 GW of new utility solar will be required, plus 24 GWh of new Battery storage, with 15 GW of new firm-dispatchable generation required with an expected utilisation of 5%. The total capital cost for this system over the next twenty-five years would be 145 billion USD based on the US EIA 2023 cost estimates (US EIA, 2023). The costs for this alternate renewable based system are detailed in Table 2.

One of the major advantages of the new wind and solar-based system is that it saves on fuel costs. Currently, the 176 TWh of generation from the coal fleet costs approximately 2.2 billion USD per year in fuel expenses, assuming coal costs approximately 100 USD per ton (GlobalEconomy.com, 2024). Once the coal-fuelled plants are decommissioned, some fuel will still be required to run the firm-dispatchable power. Assuming a 20 USD per GJ fuel, the cost of fuel for the 9 TWh of generation expected from these plants (compared to the current system's 3.6 TWh) would have an additional cost of 750 million USD. Therefore, eliminating the older coal generation plants would result in a fuel savings of approximately 75%, or over 1.7 billion USD annually. The fuel cost for the firm-dispatchable generation should be kept to a minimum, only using what is needed to keep the system balanced.

Based on these calculations, it has been determined that a renewable-based generation system would require over 30 billion USD less in capital cost than the traditional business-as-usual system in addition to the expected annual fuel savings of 1.7 billion USD. Therefore, the cost of power generation from the renewable system in the next 20 years would be 25% less expensive than that of a coal-based system. The renewable system would also produce significantly less pollution and minimal greenhouse gas emissions.

The system's reduced development schedule and modular investments also make it more responsive to changes in demand forecasts. As can be seen from the range of sizes for the coal plants in South Africa of 900 MW to 4800 MW with the two new plants at the top of this size, coal plants are generally built on large sizes to take advantage of economies of scale. This is considering locations near fuel sources as well as efficiency of the plant (Jan Van Helden & Muysken, 1981). These plants also take many years to construct. The two new coal plants in South Africa took 14 years from approval to commissioning. The World Bank indicates that coal plants take 10 to 12 years from financial close to completion, while renewable energy projects have a construction time of less than 2 years (World Bank, 2023). The peaking plants in South Africa are in sizes from 120 MW up to 1300 MW consisting of combustion engine plants and open cycle gas turbines (OCGT). These technologies can be built modularly as needed, and their construction times is less than three years. The 140 MW combustion engine power plant at Sasolburg was constructed in 15 month (Sasol, 2013) . Eskom constructed both the 1327 MW Ankerlig and 740 MW Gourikwa OCGT plants within two years (Eskom, 2014). Combustion engine and fuel cell plants are usually built with units of up to 10 MW, while OCGT plants can be built with units of up to several hundred MW. Turbine generators require some de-rating when installed at altitude and high ambient temperatures as in the High Veldt. Combustion engine plants might be preferred at these conditions (Clark, 2016). Generation plants can be made with multi-units, with expected plant sizes ranging from 100 to 1500 MW. While coal plants are generally tied to their fuel source, there is not a direct need for peaking plants to be tied to a fuel source. With low capital cost, modular sizing, short construction time and not being tied to a specific location, these firm-dispatchable plants add to a flexible program.

**Program Flexibility**

As noted in the introduction, demand forecasts are one of the most significant factors in having an optimal development program. However, this is one of the most difficult parameters to estimate. Figure 7 shows the history of the demand curve in South Africa for the last fifteen years and current growth forecast. This fifteen years is the period where the country has been working on implementing the IRP program. According to the forecast from the 2010 IRP, the demand for 2022 should have been 385 TWh per year, nearly twice the actual 2022 generation of 220 TWh (Department of Energy, 2011; Pierce & le Roux, 2023). The 2010 forecast demand would have required that over 30 GW of new generation should have been added to the system by now. With Eskom unable to meet the actual demand, it is quite fortunate that the 2010 forecast was not what happened.

The actual trend has been consistently downward at 0.5% per year on average. The demand over the last fifteen years has decreased by over 10%. Even with this decrease, Eskom has not been able to supply the demand and load shedding has increased significantly. These numbers lay out the financial consequences of demand changes not being as forecast. If the forecast had been followed an additional 30 GW of baseload generation should have been installed by now, with all its additional costs. However, no additional base load except for the two new coal plants that had been in the plan prior to commencement of the IRP.  Only a minimal amount of renewable generation was constructed, 3.4 GW of wind and 2.8 GW of solar, since the IRP program started, with no associated firm-dispatchable generation. This has not offset the loss of over 8 GW of base load generation due to performance degradation from the installed fleet. The cost of this underbuilding has been increased load shedding, with the associated indirect costs.

While no plan can withstand a "no action" plan, a development based on wind, solar, batteries and firm-dispatchable generation offers the best chance to recover from past inaction and to meet whatever the actual demand will be. The forecast for demand should be a range that commences from the current downward trend, to what is in the current IRP. The action plan should be developing a program that can meet any demand within this range. With the short two-year development times for wind, solar, batteries and firm-dispatchable generation, plus their modularity, this flexibility is reasonable.

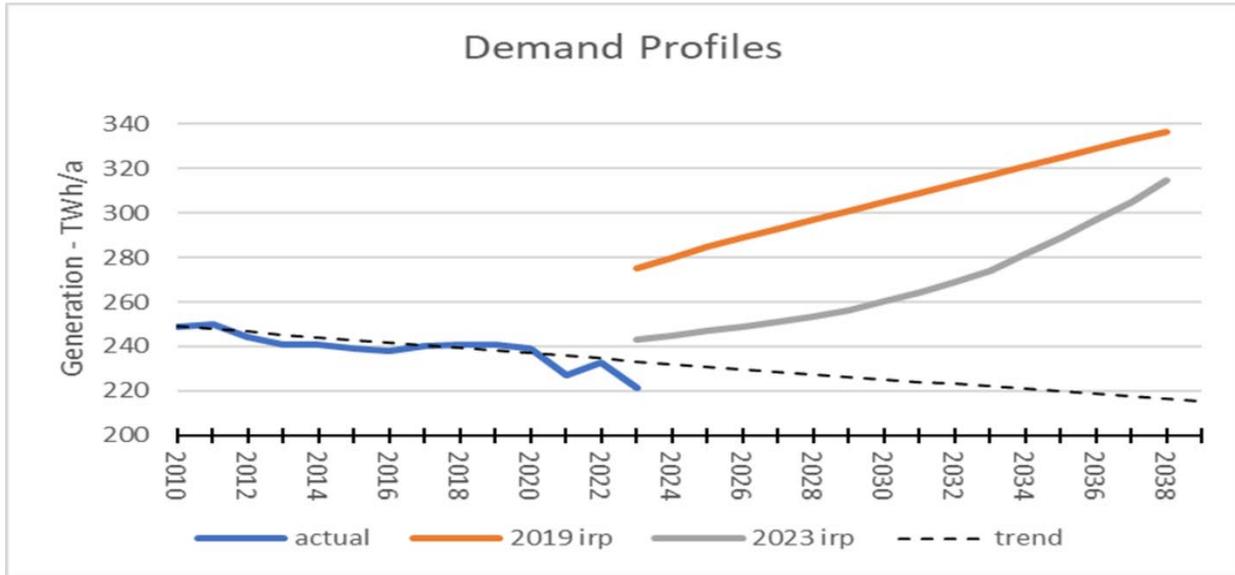

Figure 7 - South Africa electricity demand profiles – data (Pierce & le Roux, 2023; Department of Mineral Resources and Energy, 2024)

**Firm-dispatchable Generation Program**

It is anticipated that the demand for electricity generation will change through the 25-year planning cycle, and the economics of operating coal plants may affect the decommissioning program. However, it is essential to plan the development of firm-dispatchable generation, with a target of constructing 15 GW of new capacity over the next 25 years, or approximately 0.75 GW per year. This will cost around 650 million USD annually, less than 10% of the total investment program. Immediate implementation of this program is necessary to meet the current impact of load shedding which is anticipated to last through 2027 according to the latest IRP (Department of Mineral Resources and Energy, 2024). This addition of generation should be done to keep the total Eskom firm capacity at a minimum of 35 GW assuming that the demand remains at the current level.

In a report in 2018, the CSIR calculated that the immediate need for additional firm-dispatchable power was over 5 GW at that time (Wright, Calitz & Kamera, 2018). As the base load generation fleet performance has continued to deteriorate since that analysis, the current shortfall for firm-dispatchable generation is over 10 GW. After the current shortfall is overcome, the implementation schedule can be adjusted as changes arise in the demand profile, decommissioning program, and wind and solar energy implementation.

Firm-dispatchable power generation requires specific technology performance parameters, which include fast ramp rates. This can be achieved with internal combustion engine (ICE) plants, gas turbines (OCGT or CCGT), and fuel cells. Currently, both combustion engine and OCGT plants are used in the country. Fuel cells are more expensive, but they have higher efficiency. With further development, the technology is expected to eventually become economically competitive. These technologies can use various fuels, including natural gas, LPG, diesel, biofuel, and clean hydrogen. The choice of fuel is generally based on the delivered cost, which must include the logistics cost of delivery to the plant and the seasonal storage costs to keep fuel volumes sufficient to run the plant for several days. With the range of potential fuels and the ability to deliver them to most locations, the siting for firm-dispatchable generation is not tied to its fuel source as much as coal-fuelled plants, typically built at coal mine locations to minimise fuel transport. Due to its convenience in usage from easy delivery and storage, Eskom has traditionally used diesel for its peaking plants. However, diesel is an expensive choice and the budget for this fuel has become a major issue for Eskom. Going forward, Eskom is likely to switch to alternate fuels where-ever possible to reduce costs. The choices could be natural gas, LPG or biofuels.

In South Africa, natural gas is not an ideal choice for dispatchable power due to the lack of local production and limited distribution infrastructure, unlike in Europe or North America (Clark et al., 2022). Currently, natural gas is available through the Rompco pipeline in eastern part of South Africa. However, the gas fields supplying the Rompco pipeline are approaching the end of their life, making it difficult to plan around this supply (Creamer, 2019). Natural gas importation facilities in the form of LNG terminals or pipeline gas are quite capital-intensive. Natural gas would only become a reasonable alternative if other gas markets developed to support its importation, which is considered unlikely by several studies (Clark et al., 2022). Storage of natural gas at the location of the generation plant is also a challenge that must be addressed.

LPG has been recommended as an alternative fuel for Ankerlig and might be a reasonable choice for most plant locations (Clark, McGregor & Van Niekerk, 2022). As this fuel historically has been about 60% of the price of diesel fuel on the international market, it could be an economically attractive choice. This fuel does have the convenience of being a liquid fuel like diesel at nominal pressure and temperature, with related convenience of delivery and storage. This is a fuel that could be immediately used and will likely be less expensive to import and use than LNG will be.

The choice of fuel can change over time. As the use of these plants is intended to be minimised, greenhouse gas emissions from these plants are a lower priority for most systems. Over time, these plants must shift to zero-emission fuels such as hydrogen, ammonia, or biofuel.

**Locations**

Firm-dispatchable generation can be installed in locations that maximise its value within the system. This generally implies siting these plants close to major load centres and in places where the grid can handle the input without major changes.

The location of firm-dispatchable power generation should be chosen based on where it can provide the most value to the power grid at the lowest cost. This can be based on economic or social factors, among other considerations. The "Just Transition" concept is a part of government policy that aims to minimize disruptions to current employment patterns during the energy transition (Govt of South Africa, 2023). Therefore, it is important to consider this concept when selecting the location for firm-dispatchable power generation.

One possible solution for placing firm-dispatchable generation plants that meet social and economic parameters for maximum value is to construct them at the sites of decommissioned coal-fired power plants. This can be achieved by removing the base load generation plant while retaining the substation and grid connection. The new firm-dispatchable generation plant can then be built at the former base load plant location, in a size that is deemed suitable.

From a grid integration perspective these sites are expected to have a mostly positive impact on the grid and minimal negative consequences. The infrastructure already in place at these locations makes it easy to integrate power into the grid, allowing for firm generation replacement up to the capacity of the decommissioned coal plant. As a result, there should be no doubt about the grid's ability to handle this power input, given the existing grid connection.

From an economic perspective, these locations offer two advantages: minimized grid connection costs due to the use of existing connections, and construction on an existing site. One of the major costs and time factors in building any power plant is in the permitting process. Environmental and social parameters must be studied in detail for any new construction. Building on the site of an existing operation will minimize this cost and time. Eliminating the coal plant will have a positive impact on the environment by reducing pollution and improving the health of the community. However, the use of any new fossil fuel in the generation of dispatchable power may still result in some greenhouse gas and pollution emissions. These emissions will be far less than those produced by the original coal-fuelled power station. With eventual conversion of the fuel to green hydrogen, derived from excess renewable generation, these emissions can be eliminated. To maximize the value of the firm-dispatchable generation and eliminate the emissions from the plant, it should be feasible to co-locate a hydrogen production and storage on the site of the former base load plant. This should maximize the value of the plant for the net-zero grid.

From the perspective of the Just Transition, repurposing former baseload power plant sites to firm-dispatchable power can help minimize the impact on local employment. The skill sets required to operate the new plant are similar to those used in the operation of the coal power station, requiring only minimal retraining. Employment related to the supply of coal to the plant will be eliminated with the decommissioning of the existing plant in any scenario. However, there will be some employment opportunities created with the fuel supply for the new plant. While decommissioning any major baseload plant will cause significant employment disruptions, using the site for some of the required firm-dispatchable generation can help reduce this impact.

**Comparison Case**

South Africa is not the only location that is going through the energy transition. Many utility operators are considering the repurposing of baseload plants as firm-dispatchable generation to facilitate the energy transition (World Bank Group, 2021; Henkel, 2023). One recent example is the rebuilding of a 95-year-old power plant in the US state of Florida, renamed the "Dania Beach Clean Energy Center". Florida Power and Light, the plant's operator, plans to rebuild it as a firm-dispatchable plant with two OCGTs with each generating 580 MW. The operator has stated that this repurposed plant will become a major factor in support of the rapid deployment of solar generation in their system (Florida Power and Light, n.d.). They have also indicated that the plant can be converted to hydrogen fuel when it becomes economically viable.

The Florida plant will be upgraded by completely replacing the old plant with a new one at the same location. A photograph of the original plant built in 1927 is shown in Figure 8. The new firm-dispatchable energy plant will replace a base load gas plant that was built at the same location in the 1990s. Computer renderings from the operator of the existing gas plant and the new plant can be seen in Figures 9 and 10 (Florida Power and Light, n.d.). This approach can also be implemented at the baseload coal power stations in the Eskom system.

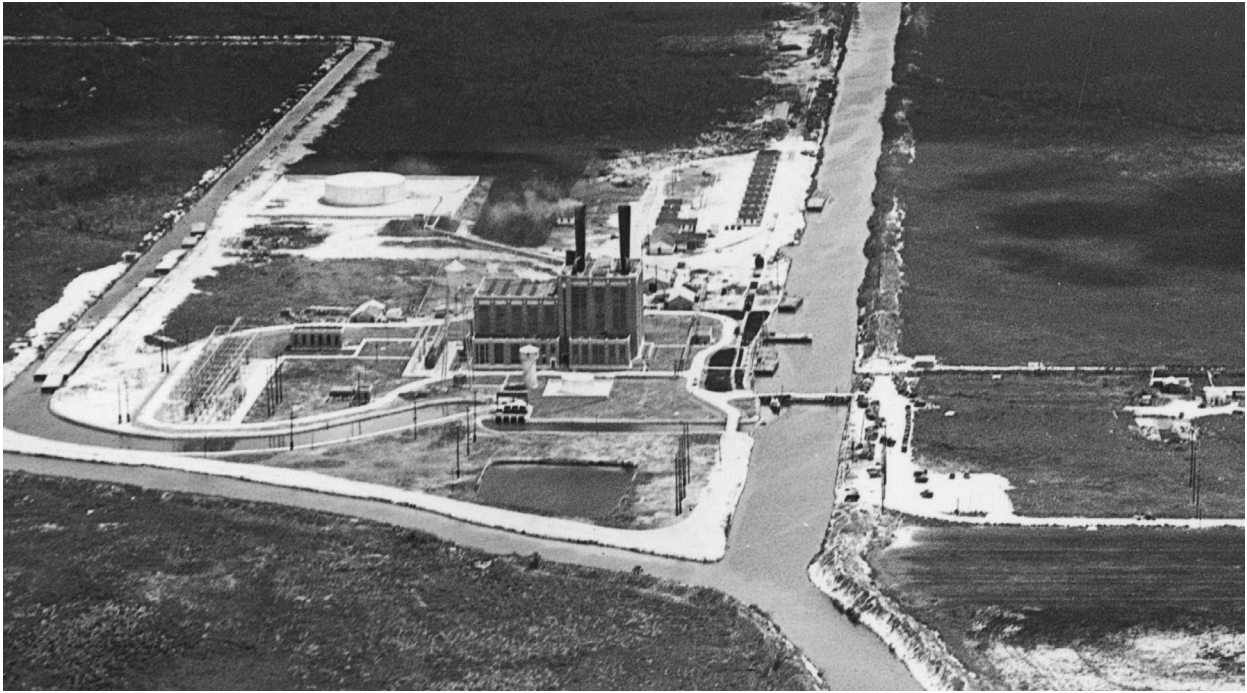

*Figure 8 - Dania Beach Power Plant 1927* (Florida Power and Light, n.d.)

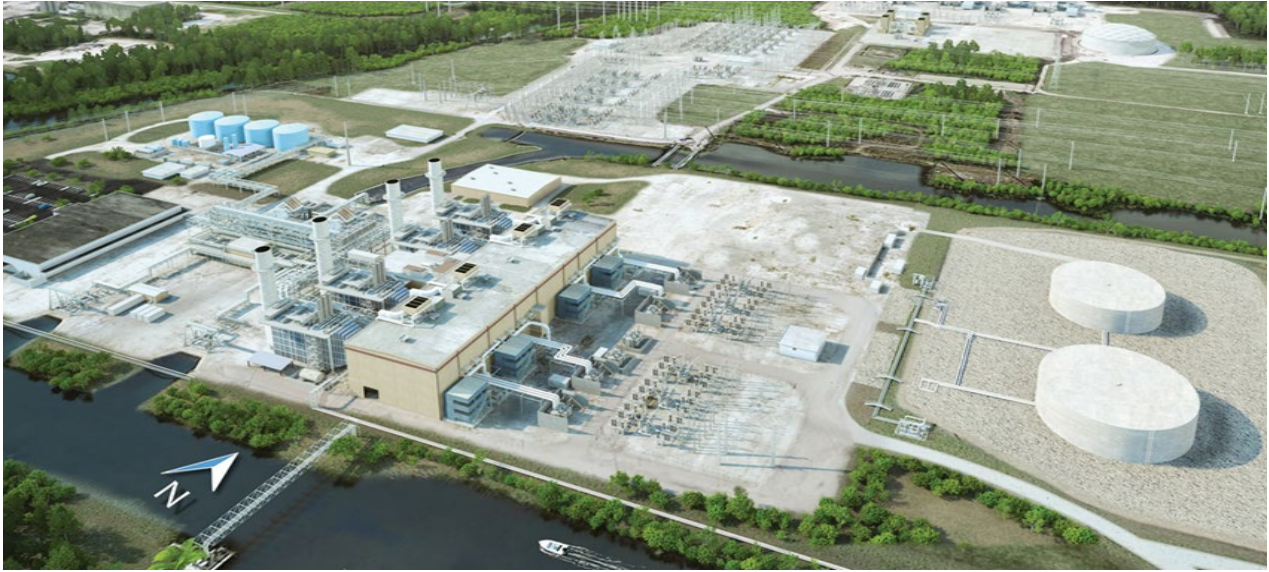

*Figure 9 - Computer Rendering of the current Dania Beach Power Plant* (Florida Power and Light, n.d.)

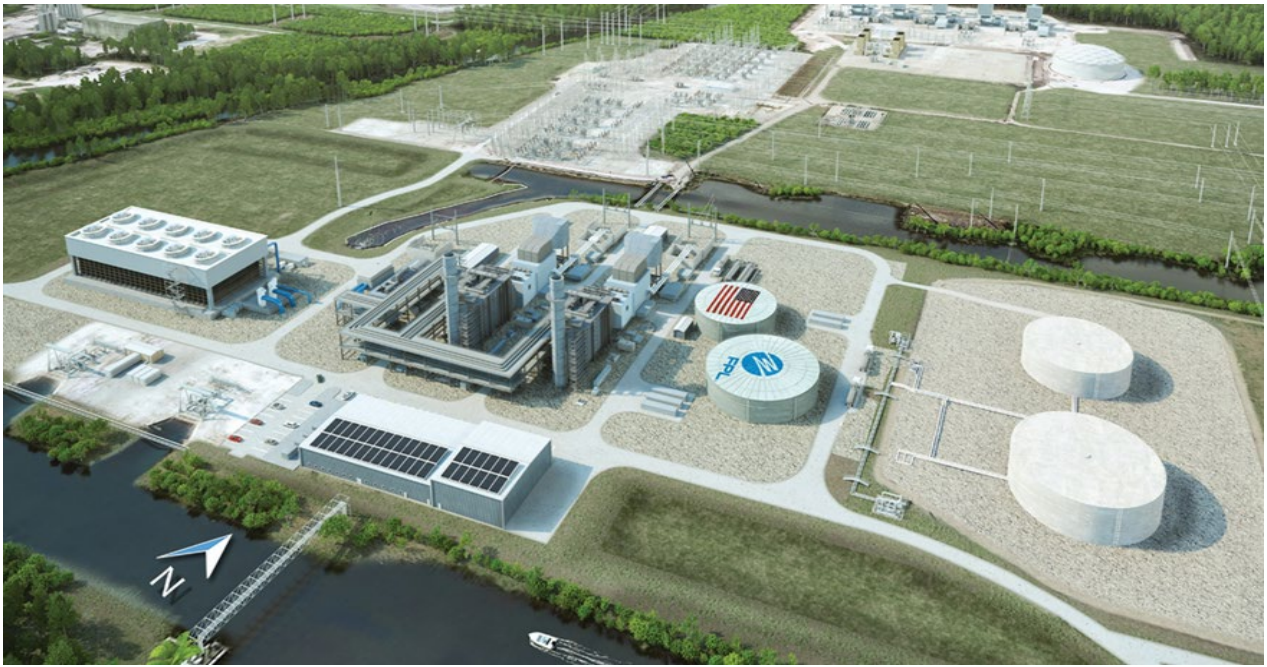

*Figure 10 - Computer Rendering of the Proposed Dania Beach Clean Energy Plant* (Florida Power and Light, n.d.)

**Implementing the Firm-dispatchable Generation Program**

The Climate Council has recommended a transition program to develop wind and solar resources (Creamer, 2023). However, this program should also include the installation of firm-dispatchable generation (Clark & McGregor, 2024). It is recommended that reputable private investors be involved in this program to meet the required deployment at a reasonable cost for the South African public. This program should be structured properly, consistent with the best practices the REIPPP Program, and commence immediately to contribute to load-shedding elimination in the short-term but also in a manner that is consistent with long-term needs and the Just Transition program.

The steps in this implantation can generally be described as follows.

1. Set up Investment Model

    The first step is to develop the financial and bidding model for implementing a firm-dispatchable generation program. Private investment must be encouraged to build these plants with the understanding that they are intended to be utilized as little as possible. This would most likely indicate a capacity payment system for guaranteed available power plus a use fee that would cover the fuel and variable operating and maintenance costs associated with the plants use. The primary payback from the system should be associated with having the capacity available rather than its use.

2. General Engineering of the Siting and Schedule of the developments

    The second step would be the general engineering and siting analysis for the base program as well as the development for a priority development schedule. It is clear from the Eskom coal plant decommissioning program that three of the plants are currently due, or overdue, for decommissioning. The first step should be the repurposing of these sites by replacing the plant. The remaining plant replacements will occur as the other plants are decommissioned. While the locations of the existing plants will provide location for some of the required firm-dispatchable generation, there will be other locations near load centres that also offer advantages that must be considered.

3. Bidding for developments

    With the overall financial model developed, the bidding for specific plant developments would be the next step in the process. The successful REIPPP program would be a reasonable starting point for the development of the bidding for these plants.

4. Installation and operation of plants

    Once the bidding process is in place, it would be up to the private investors to build out the required plants. This would likely involve completing the approval process, the removal of the existing plant, the installation and commissioning of the new plant as well as the operation of the plant once commissioned.

5. Monitoring of the need for modifications to the schedule and implementation

    One of the advantages of this program compared to the development program for baseload generation is the ability to adapt the program to meet changing need. A monitoring program should be in place to adapt the program as required during its implementation.

6. Engineering analysis of conversion to zero-emission fuels

    Eventually, these plants must be part of a net zero generation system and as such, there must be an engineering program to make the conversions of these plants to zero emission fuels. This could take the form of hydrogen generated from the excess renewable generation or potentially from biofuels. Neither of these conversions should be technically difficult and should not affect the development or use of these facilities.

**Conclusion**

In conclusion, this research underscores the critical need for expediting a firm-dispatchable generation program within South Africa's evolving energy landscape. The imminent decommissioning of ageing coal-fired plants, coupled with the uncertain future demand for energy as well as the nation's commitment to a Just Transition, necessitates a strategic shift towards renewable sources supported by firm-dispatchable power. The proposed 25-year plan, emphasising modular technologies and repurposing decommissioned coal plant sites, offers a cost-effective and environmentally sound alternative to the traditional rigid coal-centric paradigm. The integration of private investors, drawing inspiration from global case studies, and adherence to established best practices enhance the feasibility and adaptability of the recommended approach. The envisaged program addresses immediate challenges and lays the foundation for a sustainable and resilient energy infrastructure that aligns with global imperatives and the principles of a Just Transition.